\documentclass[aps,prl,twocolumn,superscriptaddress,showpacs]{revtex4-2}
\usepackage[sort&compress]{natbib}
\usepackage[utf8]{inputenc}
\usepackage{graphicx}
\usepackage{dcolumn}
\usepackage{braket}
\usepackage{setspace}
\usepackage{bm}
\usepackage{amsmath}
\usepackage{csquotes}
\usepackage{mathrsfs}
\usepackage{titlesec}
\usepackage{subfigure}
\usepackage[svgnames]{xcolor}
\usepackage[colorlinks,citecolor=red,urlcolor=red,bookmarks=false,hypertexnames=true]{hyperref}
\usepackage{soul}

\newcommand{\pare}[1]{\left( #1 \right)}
\newcommand{\llav}[1]{\left\lbrace #1 \right\rbrace}

\begin{document}

\title{High-dimensional encryption in optical fibers using machine learning}

\author{Michelle L. J. Lollie}
\thanks{These authors contributed equally to this work.}
\affiliation{Quantum Photonics Laboratory, Department of Physics \& Astronomy, Louisiana State University, Baton Rouge, LA 70803, USA}

\author{Fatemeh Mostafavi}
\thanks{These authors contributed equally to this work.}
\affiliation{Quantum Photonics Laboratory, Department of Physics \& Astronomy, Louisiana State University, Baton Rouge, LA 70803, USA}

\author{Narayan Bhusal}
\affiliation{Quantum Photonics Laboratory, Department of Physics \& Astronomy, Louisiana State University, Baton Rouge, LA 70803, USA}

\author{Mingyuan Hong}
\affiliation{Quantum Photonics Laboratory, Department of Physics \& Astronomy, Louisiana State University, Baton Rouge, LA 70803, USA}

\author{\mbox{Chenglong You}}
\affiliation{Quantum Photonics Laboratory, Department of Physics \& Astronomy, Louisiana State University, Baton Rouge, LA 70803, USA}

\author{Roberto de J. Le\'{o}n-Montiel}
\affiliation{Instituto de Ciencias Nucleares, Universidad Nacional Autónoma de México, Apartado Postal 70-543, 04510 Ciudad de México, México}

\author{Omar S. Magaña-Loaiza}
\email{Corresponding Author: maganaloaiza@lsu.edu}
\affiliation{Quantum Photonics Laboratory, Department of Physics \& Astronomy, Louisiana State University, Baton Rouge, LA 70803, USA}

\author{Mario A. Quiroz-Juárez}
\affiliation{Departamento de Física, Universidad Autónoma Metropolitana Unidad Iztapalapa, San Rafael Atlixco 186, 09340 Ciudad México, México}

\date{\today}

\begin{abstract}
The ability to engineer the spatial wavefunction of photons has enabled a variety of quantum protocols for communication, sensing, and information processing. These protocols exploit the high dimensionality of structured light enabling the encodinng of multiple bits of information in a single photon, the measurement of small physical parameters, and the achievement of unprecedented levels of security in schemes for cryptography. Unfortunately, the potential of structured light has been restrained to free-space platforms in which the spatial profile of photons is preserved. Here, we make an important step forward to using structured light for fiber optical communication. We introduce a smart high-dimensional encryption protocol in which the propagation of spatial modes in multimode fibers is used as a natural mechanism for encryption.  This provides a secure communication channel for data transmission. The information encoded in spatial modes is retrieved using artificial neural networks, which are trained from the intensity distributions of experimentally detected spatial modes. Our on-fiber communication platform allows us to use spatial modes of light for high-dimensional bit-by-bit and byte-by-byte encoding. This protocol enables one to recover messages and images with almost perfect accuracy. Our smart protocol for high-dimensional optical encryption in optical fibers has key implications for quantum technologies relying on structured fields of light, particularly those that are challenged by free-space propagation. 

\end{abstract}

\maketitle
Among the multiple families of structured optical beams, Laguerre-Gaussian (LG) modes have received particular attention for their ability to carry orbital angular momentum (OAM) \cite{allen1999iv,AllenOAM_PRA,yao:2011,padgett2017orbital}. The OAM in this kind of beam is induced by a helical phase front given by an azimuthal phase dependence of the form $e^{i\ell\phi}$, where $\ell$ represents the OAM number and $\phi$ the azimuthal angle \cite{allen1999iv,AllenOAM_PRA,yao:2011,padgett2017orbital,omar:2019}. Over the past decade, there has been an enormous interest in using photons carrying OAM for quantum communication \cite{willner:15, WillnerLiu:2021, wang:2012, baghdady:2016, Liu:2019, Cozzolino:2019,yan2014high}. These structured beams of light allow for the encoding of multiple bits of information in a single photon \cite{yao:2011, padgett2017orbital, omar:2019}. Additionally, it has been shown that high-dimensional Hilbert spaces defined in the OAM basis can increase the robustness of secure protocols for quantum communication \cite{mirho:2015,Rodenburg_2014, Forbes:2021}. However, despite the enormous potential of structured photons, their vulnerabilities to phase distortions impose important limitations on the realistic implementation of quantum technologies \cite{malik:2012, bhusal2021spatial,willner:15, omar:2019, yao:2011, milione:2017, lavery:2013, Magana2016hanbury, mirho:2015, malik:2012, willner:15}. Indeed, LG beams are not eigenmodes of commercial optical fibers and consequently their spatial profile is not preserved upon propagation. For this reason,  quantum communication with structured photons has been limited to free-space platforms \cite{gibson2004free,anguita2008turbulence,wang2012terabit,ren2016experimental, su2012demonstration, xie2015performance}.

Recently, there has been an enormous interest in employing artificial neural networks to boost the functionality and robustness of quantum technologies  \cite{krenn2020computer, Beer:2020, bishop2006pattern, lecun2015deep, carleo2019machine, Walln:2020}. In the field of photonics, there has been extensive research devoted to developing artificial neural networks for the implementation of novel optical instruments \cite{you2020identification, kudyshev2020rapid, gebhart2020neural}. Indeed, convolutional neural networks (CNNs) have enabled the demonstration of new imaging schemes working at the single-photon level \cite{giordani2020machine, doster2017machine, bhusal2021spatial}. These protocols have been employed to characterize structured photons in the Laguerre-Gaussian (LG), Hermite-Gaussian (HG), and Bessel-Gaussian (BG) bases \cite{doster2017machine, sun2019identifying, giordani2020machine, hofer2019hermite, park2018multiplexing, willner:15, omar:2019, yao:2011, Huang:2021}. Here, we introduce a machine learning protocol that exploits spatial modes of light propagating in multimode fibers for high-dimensional encryption. This is achieved by training artificial neural networks from experimental spatial profiles in combination with a theoretical model that describes the propagation of spatial modes in multimode fibers. The trained neural network enables us to decrypt information encoded in spatial modes of light. We demonstrate robust and efficient bit-by-bit and byte-by-byte encryption in commercial multimode fibers.

\begin{figure*}
\centering  
\includegraphics[width=0.95\linewidth]{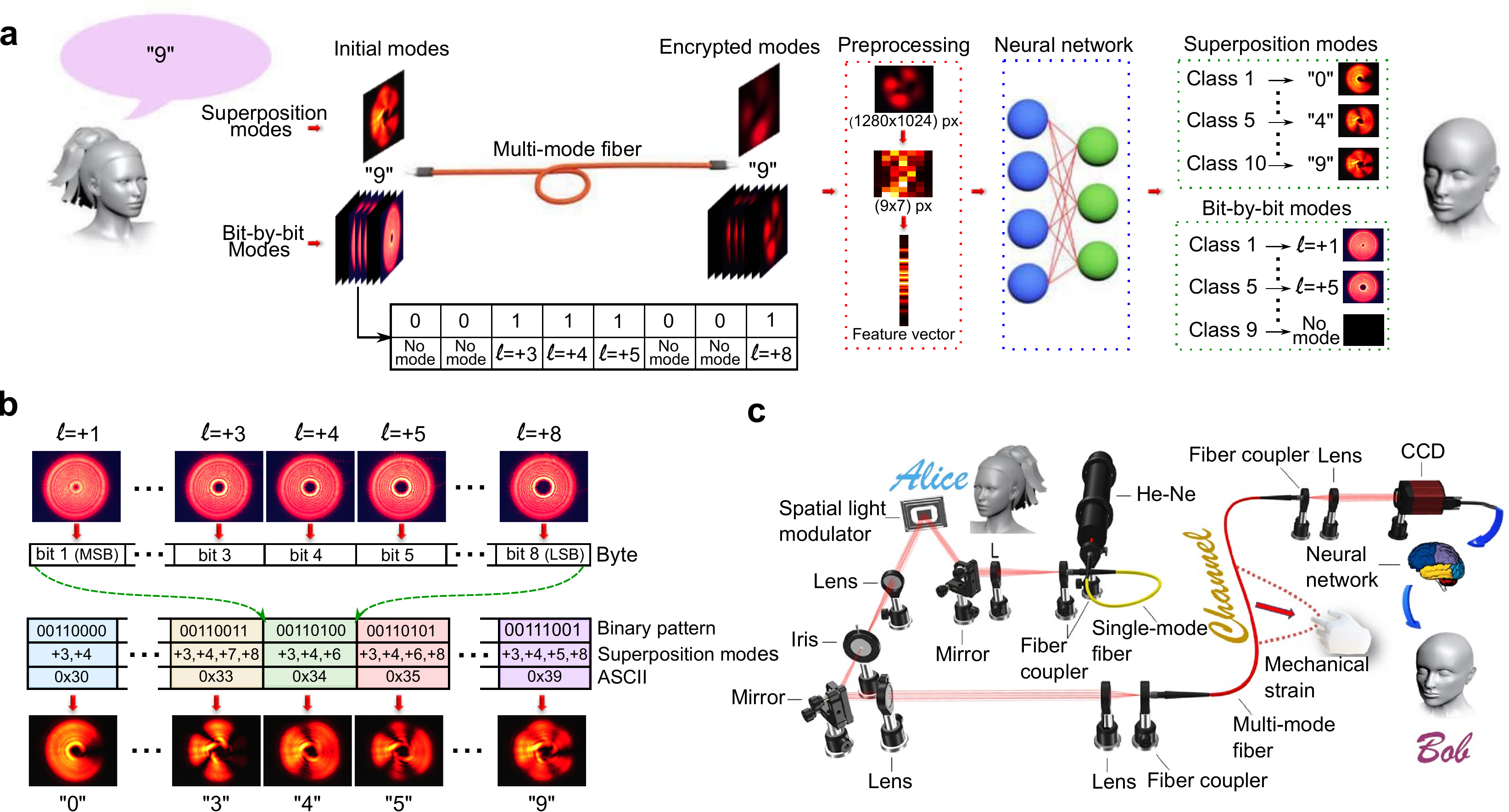}
%\subfigure[]{\includegraphics[width=1\linewidth]{setupconcept.jpg}}
%\newline
%\subfigure[]{\includegraphics[width=1\linewidth]{finalsetup.jpg}}
\caption{\textbf{a} Conceptual schematic of our encryption protocol. In this case, Alice sends the message \enquote{9} to Bob in a bit-by-bit fashion or through a superposition of spatial modes (byte-by-byte). The resulting computationally efficient feature vector is used to train a neural network with high accuracy. The preprocessing details for encrypted modes can be found in the Methods section. \textbf{b} The OAM mode-to-bit-position relation is shown along with superposition states that correspond to the ASCII digits from zero to nine. This \enquote{alphabet} is used to encode information in spatial modes carrying OAM. Our experimental setup is depicted in \textbf{c}.  Here, Alice encodes a message using OAM modes generated through a spatial light modulator (SLM). The spatial modes are coupled into a multimode fiber that is used to transmit information to Bob. In this case, we emulate multiple transmission conditions by introducing stress to the fiber via mechanical manipulation. The resulting perturbations are used to encrypt the message. We train our artificial neural network by collecting multiple spatial profiles of the distorted beams produced by the multimode fiber. Each distorted spatial profile of the optical beam corresponds to a particular condition of stress exerted on the fiber. Remarkably, our neural network is capable of recovering the initial superposition modes, converting them to the standard alphabet characters for Bob at read out.}
\label{fig:concept}
\end{figure*}

The conceptual illustration of our smart encryption protocol is presented in Fig. \ref{fig:concept}a. Here, Alice prepares a message encoded in high-dimensional OAM modes that is then sent to Bob through a multimode fiber. The protocol entails the use of the 8-bit ASCII (American Standard Code for Information Interchange) code, allowing Alice to encode a message in two different ways using the alphabet shown in Fig. \ref{fig:concept}b. In the first approach, each character in the message is represented by a byte (eight bits). Then, the OAM modes from $\ell=+1$ to $\ell=+8$ are one-to-one correlated with the position of each bit. The mode $\ell=+1$ is the most significant bit (MSB) and $\ell=+8$ the least significant bit (LSB). Consequently, for each character in the message, Alice has an eight-bit binary string where each bit position is mapped to an LG mode and sequentially sent to Bob in a bit-by-bit fashion. In the second approach,  Alice prepares one state (byte) composed of a superposition of eight bits representing a particular character. This enables Alice to send a message to Bob through a sequence of characters, or byte-by-byte, leading to a more computationally efficient process. This scheme enables the mitigation of some errors that may be introduced during the transmission and reception of information, such as the loss of bits. 

We decrypt messages by training artificial neural networks with experimental spatial profiles in combination with a theoretical model that describes our protocol. We now introduce our model to describe propagation of spatial modes in multimode fibers. For this purpose, we consider the coupling of an encoded message from free space to the transmission channel, namely the optical fiber. In this case, one has to decompose the initially injected field into the modes that are sustained by the specific features of the fiber. For the weakly guiding step-index fiber used in this experiment, the modes are described by the linearly polarized (LP) solution set. The field distribution, in polar coordinates, $\Psi\pare{r,\phi}$, is thus described by the solution of the scalar Helmholtz equation, which for a cylindrical fiber with a core radius $a$ is given by \cite{saleh_book,fibers_book,brunin2015}
\begin{equation}
\text{LP}_{\ell p} = N_{\ell p}
\left\{
	\begin{array}{ll}
		\text{J}_{\ell}\pare{\kappa_{T\ell p}r}\exp\pare{-i\ell \phi}  & \mbox{if } r < a, \\
		\text{K}_{\ell}\pare{\gamma_{\ell p} r}\exp\pare{-i\ell \phi} & \mbox{if } r \geq a,
	\end{array}
\right.
\end{equation}
where $N_{\ell p}$ is a normalization constant, $J_{\ell}\pare{x}$ is the Bessel function of the first kind and order $\ell$, and $K_{\ell}\pare{x}$ is the modified Bessel function of the second kind and order $\ell$. Note that the parameters $\kappa_{T\ell p}$ and $\gamma_{\ell p}$ determine the oscillation rate of the field in the core and the cladding, respectively. These are defined by
\begin{eqnarray}
\kappa_{T\ell p}^{2} &=& n_{\text{core}}^{2}k_{0}^{2} - \beta_{\ell p}^{2}, \\
\gamma^{2}_{\ell p} &=& \beta_{\ell p}^{2} - n_{\text{cladding}}^{2}k_{0}^{2},
\end{eqnarray}
where $k_{0}=2\pi/\lambda_{0}$, with $\lambda_{0}$ being the vacuum wavelength of the light inside the fiber, $\beta_{\ell p}$ is the propagation constant of the $p$th guided mode for each azimuthal index $\ell$, and $n_{\text{core}}$ and $n_{\text{cladding}}$ are the refractive indices of the core and the cladding, respectively. For the description of the LP modes, the additional fiber parameter $V$ is required, which is defined as
\begin{equation}
V^{2} = \kappa_{T\ell p}^{2} + \gamma_{\ell p}^{2} = \pare{2\pi\frac{a}{\lambda_0}}^{2}\pare{n_{\text{core}}^{2} - n_{\text{cladding}}^{2}}.
\end{equation}
This fiber parameter determines the amount of modes and their propagation constants. In our experiments, we make use of a 1-meter long, 10 $\mu m$-diameter fiber, with $\text{N.A.} = \sqrt{n_{\text{core}}^{2} - n_{\text{cladding}}^{2}} = 0.1$. In these conditions, an arbitrary field propagating along the fiber may be decomposed in six LP modes with indexes $(\ell,p) \in \llav{(-2,1),(-1,1),(0,1),(1,1),(2,1),(0,2)}$. This implies that, regardless of the initial condition, the output mode of the fiber can always be written as
\begin{equation}\label{eq:coeff}
\Psi_{\text{out}}\pare{r,\phi} = \sum_{\ell,p}c_{\ell,p}\text{LP}_{\ell,p},
\end{equation}
where the coefficients $c_{\ell,p}$ are defined by the injected field and the properties of the optical fiber throughout the propagation length.

%\begin{figure}[t!]
%\centering
%\includegraphics[width=8cm]{Modes-eps-converted-to.pdf}
%\caption{Modes sustained by the 1-meter long, $\text{N.A.} = 0.1$, 10 $\mu m$-diameter optical fiber used in our experiments.}
%\label{fig:modesfiber}
%\end{figure}

In a realistic scenario, the local random variations of the fiber properties produce significant distortions of spatial modes, thus making almost impossible to predict the spatial distribution of photons at the end of the fiber, i.e., the coefficients $c_{\ell,p}$ in Eq. (\ref{eq:coeff}). This is the main motivation behind our machine-learning protocol for  encryption in optical fibers.
%which can be exploited to get an encrypted message. 
In our experiments, variations are produced by a mechanical strain,
%Aberrations induced by the fiber produce significant distortions of the optical beams, which are exploited to get an encrypted message. In our experiment, these aberrations are emulated by a mechanical strain, 
and in the case of superposition modes, by both strain and mixing of the modes during the propagation. Once optical spatial modes leave the fiber, the goal is to recover the sequence of transmitted modes either bit-by-bit (individual modes) or byte-by-byte (superposition modes) as the case may be and then effectively decode the optical profiles (images) to compose the message. In this respect, Bob exploits the self-learning features of artificial neural networks to decrypt the information encoded in the distorted spatial modes efficiently. To train the neural network, the data-set comprises a collection of down-sampled images, rearranged as column vectors that correspond to the aberrated optical profiles, as shown in Fig. \ref{fig:concept}a. After the training, Bob utilizes the high efficiency of the neural network to retrieve the message by identifying individual modes if the communication was bit-by-bit, or recognising superposition modes when the communication was byte-by-byte.

\begin{figure}[t!]
\centering  
\includegraphics[width=0.95\linewidth]{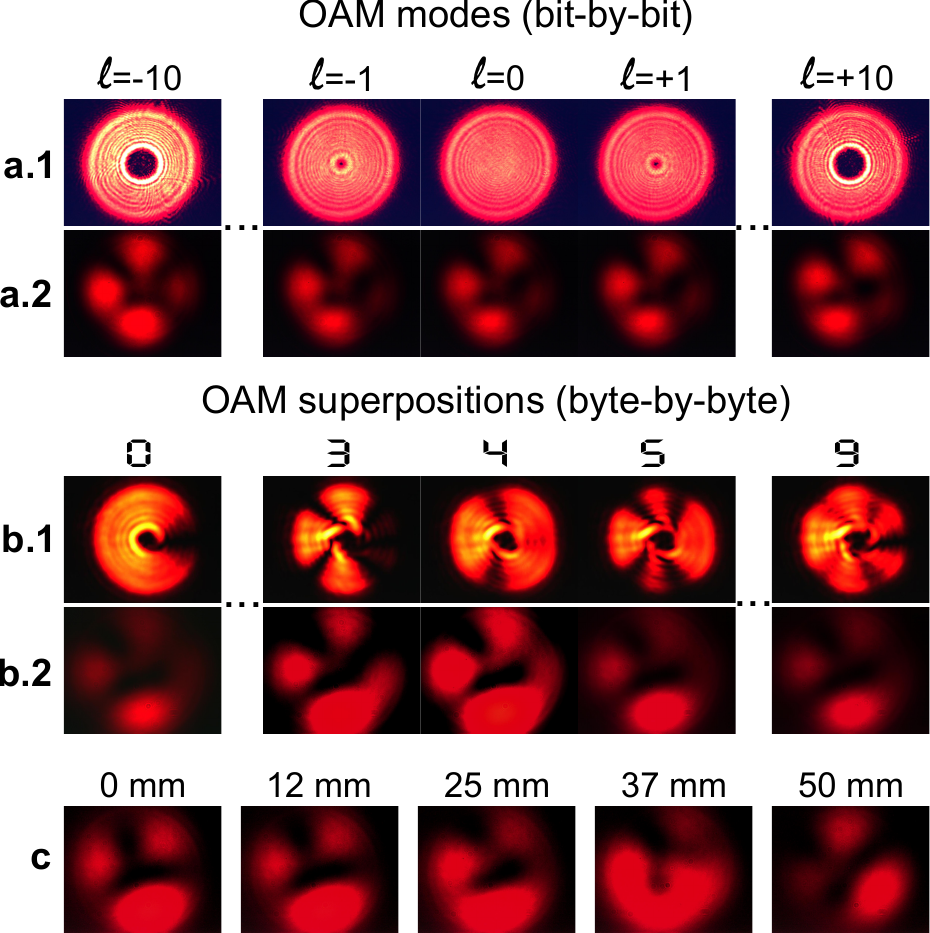}
\caption{Spatial intensity distributions of initial and encrypted LG modes obtained experimentally for the maximum strain in the fiber (50 mm).  Intensity profiles of individual modes with azimuthal quantum numbers, $\ell=-10$, $\ell=-1$, $\ell=0$, $\ell=+1$, $\ell=+10$, before (\textbf{a.1}) and after (\textbf{a.2}) the propagation through the multimode fiber. \textbf{b.1} Superposition of LG modes representing the numeric characters 0, 3, 4, 5, and 9. Each character has been encoded using the alphabet displayed in Fig. \ref{fig:concept}. The bottom row (\textbf{b.2}) shows the encrypted modes corresponding to each of the superpositions. \textbf{c} Spatial profiles for the numeric character 1 obtained after propagation for different displacements of the fiber: 5, 12, 25, 37, and 50mm. Note that the fiber experiences strain due to the displacement, resulting in a dynamic intensity output.}
\label{fig:modes}
\end{figure}

%``1''
The schematic diagram of our experimental setup is shown in Fig. \ref{fig:concept}c. Alice use a He-Ne laser with a spatial light modulator (SLM) to prepare the message to be sent using OAM states of light. The light beam is then sent to Bob through multimode fiber, and the output is measured by a camera. More details of the experiment can be found in the Methods section. Examples of different OAM intensity distributions collected experimentally are shown in Fig. \ref{fig:modes}. Spatial profiles of individual LG modes with different azimuthal quantum numbers ($\ell=-10$, $\ell=-1$, $\ell=0$, $\ell=+1$, $\ell=+10$) and LG superpositions, before the multimode fiber, are displayed in Figs. \ref{fig:modes}a.1 and b.1, respectively. Each superposition has unique intensity distribution given by combining two and up to eight OAM single modes, depending on the character to be represented (see the alphabet shown in Fig. \ref{fig:concept}b). For demonstration purposes, Fig. \ref{fig:modes}b.1 presents superpositions of LG modes for the numeric characters: 0, 3, 4, 5 and 9.

\begin{figure*}
\centering  
\includegraphics[width=0.95\linewidth]{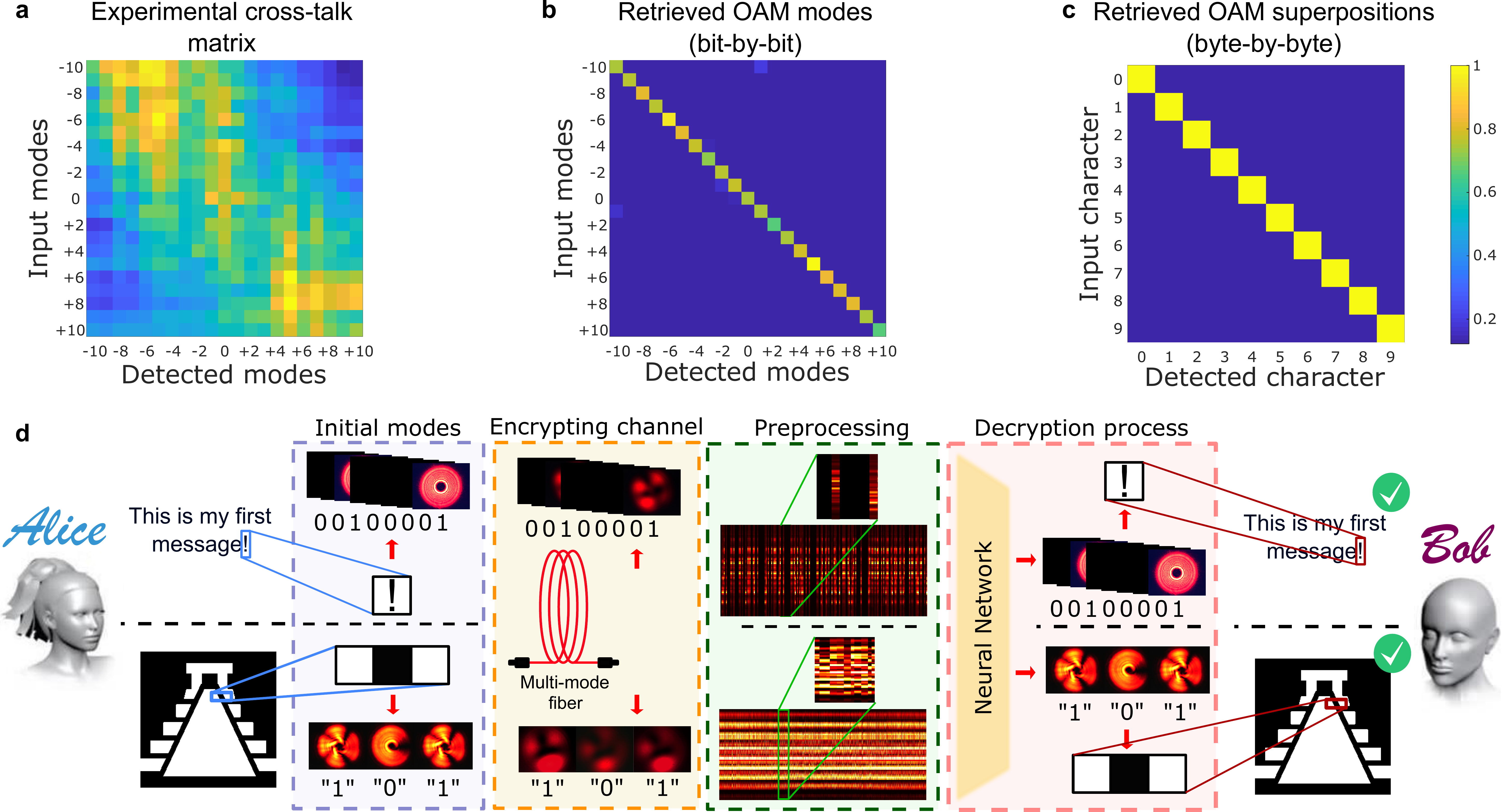}
\caption{\textbf{a} Cross-talk matrix for LG modes with azimuthal quantum numbers from $\ell=-10$ to $\ell=+10$. Note that the LG modes are distorted severely, so the identification is practically impossible. \textbf{b} Diagonal cross-talk matrix obtained after applying our neural network. Our approach provides a powerful tool to recognize OAM modes after the fiber with an efficiency of 98\%. \textbf{c} The cross-talk matrix obtained for the LG superpositions that represent the numeric characters zero to nine. The diagonal elements indicate that the transmitted characters are correctly identified. This neural network exhibits a performance of 99\%. \textbf{d} Smart communication protocol with Alice sending the message \enquote{This is my first message!} (upper message) and the image of a Mexican pyramid (bottom message) to Bob through the multimode fiber. The transmission of the text and image is bit-by-bit and byte-by-byte, respectively. Bob deciphers both messages by using the trained NN with near-unity accuracy.}
\label{fig:cross}
\end{figure*}

As mentioned above, once a spatial mode is transmitted through the fiber, there is significant distortion from the applied tension, the local variations of the fiber properties, and even the noise of the camera sensor,  resulting in an encrypted mode. Fig. \ref{fig:modes}a.2 and b.2 show the encrypted modes corresponding to the spatial beams in Fig. \ref{fig:modes}a.1 and b.1, respectively, for a displacement of 50 mm, which represents the maximum strain that may be applied in the fiber in our experiment. Note that intensity distributions of the encrypted modes change drastically with respect to the distributions of the initial modes. Moreover, the shape of the intensity patterns can change significantly as a function of the strain experienced by the fiber. Fig. \ref{fig:modes}c displays spatial profiles for LG superposition of the character 1 with different applied tension represented by the displacements: 0, 12, 25, 37, and 50 mm. Importantly, these distortions are induced randomly, which leads to an unbounded set of encrypted modes. Nevertheless, our NN can decode these encrypted modes with high accuracy for both individual modes and mode superpositions. This effectively generalizes an unbounded set from a limited collection of labeled examples.

To study the LG mode cross-talk when the beam propagates through the fiber, we measured the cross-correlation matrix for modes with azimuthal quantum numbers from $\ell=-10$ to $\ell=+10$. In the cross-talk matrix, the diagonal elements represent the conditional probabilities among the transmitted and detected modes that were correctly recognized. Remarkably, the diagonal elements in our cross-talk matrix are erased completely as indicated by Fig. \ref{fig:cross}a. In fact, it is practically impossible to recognize any mode. However, as shown in Figs. \ref{fig:cross}b and \ref{fig:cross}c, we exploit the functionality of machine learning algorithms to design a NN sufficiently sensitive to discern LG modes after the multimode fiber, enabling us to reconstruct a diagonal cross-talk matrix. To show the ability of our machine-learning algorithms to recognize encrypted OAM modes,  we first design, train, and test a multi-layer neural network with the capacity to identify LG beams with different positive and negative topological charges, which go from $\ell=-10$ to $\ell=+10$. It is known that two pure LG modes with identical radial numbers but with opposite topological charges are indistinguishable using intensity measurements solely because they present exactly the same distributions. However, we experimentally demonstrate that our approach enables the discrimination of oppositely charged LG modes from their intensity patterns. We exploit the fact that OAM propagation through the multimode fiber induces phase distortions. The fiber is interpreted to be a \enquote{disordered} medium due to the inherent noise and the local variations of its properties. This leads to distinct modal cross-talk for the LG modes and their conjugates, resulting in changes in the intensity profiles. Consequently, this allows the NN algorithms to distinguish opposite LG modes unequivocally. Thus, our approach overcomes the limitations of existing strategies based on projection measurements and phase-measurement interferometry techniques. As seen in Fig. \ref{fig:cross}b, we obtain a classification accuracy of 98\%.

Now we describe the implementation of the smart communication protocol using the trained NN. For bit-by-bit communication, we select the LG modes from $\ell=+1$ to $\ell=+8$ to form 8-bit binary words that allow us to encode characters from the ASCII code. It is worth mentioning that by using these eight modes, our neural network reaches an overall accuracy of 99\%. Again, Alice encodes a message using the alphabet shown in Fig. \ref{fig:concept}b. This process is presented in Fig.  \ref{fig:cross}d. Alice prepares the plain text \enquote{This is my first message!} that is transmitted to Bob through the multimode fiber. Note that we show the detailed encoding and decoding processes for a particular character. In the figure, we highlight the exclamation mark, however the same stages are applied for all the characters of the message. The communication channel acts as the encryption process, so Bob receives a sequence of indistinguishable intensity profiles. The goal is to recover the sequence of transmitted bits by Alice from the intensity distributions. Prior to the decryption process, Bob carries out image pre-processing that includes the transformation of an image from RGB to grayscale. This is followed by the down-sampling process and the rearrangement of the pixels from resulting matrices into column vectors. In the decryption process, Bob uses the neural network to decipher the message by identifying each received LG mode and translating it via the standard alphabet.

To describe the implementation of our proof-of-principle smart communication protocol for byte-by-byte communication, LG superposition modes are prepared using the alphabet in Fig. \ref{fig:concept}b. We begin by using the dataset of encrypted superposition modes to train, validate and test a neural network that maps the distorted mixtures to one of the transmitted digits. After training, the performance of our neural network is 99\%. This demonstrates the ability of our neural network to discern, with near-unity accuracy, experimental superpositions of LG modes. This is highlighted via the cross-talk matrix in Fig.   \ref{fig:cross}c. Furthermore, to unveil the utility and functionality of our smart communication protocol, Fig.  \ref{fig:cross}d presents a scheme where Alice sends the image of a Mexican pyramid to Bob through the multimode fiber. As in the previous case for the plain text, we emphasize the involved processes in the communication protocol for three particular pixels from the image. Here, each pixel of the image is represented by an eight-bit word whose decimal value is ``1'' for white pixels and ``0'' for black pixels. Alice can employ the superposition modes that represent the digits ``1''  and ``0'' to map the image and transmit it byte-by-byte (or equivalently pixel-by-pixel) through the communication channel. Thus, Bob receives one by one the encrypted pixels that comprise the image and preprocesses them. Then, Bob uses the neural network to identify the digits encoded in the superposition modes, after which he can retrieve the Mexican pyramid. At this point it is worth mentioning that, after the propagation, the image information cannot be inferred from the distorted beams. This decryption process requires the trained neural network to recover the plain image.

We quantify the integrity of the received information by calculating the mean squared error (MSE), defined by $MSE=\frac{1}{n}\langle \textbf{e} | \textbf{e} \rangle$ where $\textbf{e}=(\hat{\textbf{y}}-\textbf{y})$, . Here, $\hat{\textbf{y}}$  and $\textbf{y}$ are vectors that contain the received and transmitted bytes, respectively. The measured MSE for both the message and image is zero. This validates the robustness and high efficiency of our protocol to decode OAM modes transmitted through the multimode fiber. 

In summary, we have demonstrated a machine learning protocol that employs spatial modes of light in commercial multimode fibers for high-dimensional encryption. This protocol was implemented on a communication platform that utilizes LG modes for high-dimensional bit-by-bit and byte-by-byte encoding. The method relies on a theoretical model that exploits the training of artificial neural networks for identification of  spatial optical modes distorted by multimode fibers. This process allows for the recovery of encrypted messages and images with almost perfect accuracy. Our smart protocol for high-dimensional optical encryption in optical fibers has key implications for quantum technologies that rely on structured fields of light, especially those technologies where free-space propagation poses significant challenges.

\section*{Acknowledgements}
M.L.J.L. would like to thank Louisiana State University (LSU) for financial support via the Huel D. Perkins Fellowship, and the LSU Department of Physics \& Astronomy for supplemental support. F.M. and O.S.M.L. acknowledge support from the National Science Foundation through Grant No. CCF-1838435. M.H., C. Y. and O.S.M.L. thank the Army Research Office (ARO) for support under the grant no. W911NF-20-1-0194. R.J.L.-M. thankfully acknowledges financial support by CONACyT under the project CB-2016-01/284372 and by DGAPA-UNAM under the project UNAM-PAPIIT IN102920. 

\section*{Competing Interests}
The authors declare no conflict of interest.

\section{Methods}

\section{Experiment}
The schematic diagram of our experimental setup is shown in Fig. \ref{fig:concept}c. We use a He-Ne laser at 633 nm that is spatially filtered by a single-mode fiber (SMF). The output beam with a Gaussian profile illuminates a spatial light modulator (SLM) displaying a computer-generated hologram. The SLM together with a 4f-optical system allows us to prepare any arbitrary spatial mode of light carrying OAM. We then use a telescope to demagnify the structured beam before coupling into a multimode fiber with diameter of 10 $\mu \text{m}$. The preparation of the modes used to store the message to be sent is performed by Alice. At the output of the fiber, Bob uses a camera to measure the collimated spatial profile of the communicated modes. Mechanical stress is induced in the fiber channel to generate the neural network training palette. The fiber is configured in a loop with the base secured to the optical table. The top of the loop is secured to a 3D translation stage with displacement occurring along the y-axis (orthogonal to and away from the plane of the table). Displacing the top of the loop attached to the translation stage 50 mm produces strain in the fiber. As the fiber is being pulled taut, successive images show the dynamic change of the mode, so the output at detection is now an LG mode distorted both via the multimode-fiber beam transformation as well as the applied tension. A camera is used to detect and display the output image. Two sets of data are taken: 1) The SLM is programmed to produce holograms for each OAM mode from -10 to 10, 21 modes total. For each mode, one image is captured at 0.10 mm translation intervals for a total displacement of 50 mm producing 500 images. 2) The SLM is programmed to produce holograms of OAM superposition modes for the 8-bit ASCII characters zero to nine. Each character is represented by a superposition of two, three, four, or five OAM single modes. One image is captured per 0.25 mm displacement interval over 50 mm for a total of 200 images per each superposition mode.

\section{NN Training}

In what follows, we describe technical aspects of the neural networks developed in this work. The acquired sets of images in combination with machine learning algorithms enable the identification of distorted LG modes. This renders the originally encoded modes (message). The machine learning algorithms are characterized by solve tasks where conventional algorithms offer low performances or limited efficiencies. Typically, these solve tasks exploit a given collection of labeled examples or \enquote{past experiences} to predict the outcome for new data \cite{bishop2006pattern, Goodfellow2016}. We implement feed-forward neural networks with sigmoid neurons in the single hidden layer and softmax neurons in the output layer, to identify spatial modes transmitted through a multimode fiber. In this architecture, each neuron in a specific layer is connected to each neuron of the next layer through a synaptic weight. These synaptic weights are optimized by using the scaled conjugate gradient back-propagation algorithm \cite{moller1993scaled} in a direction that minimizes the cross-entropy \cite{shore1981properties, de2005tutorial}. Because sigmoid neurons are ranged in the interval [0,1], the cross-entropy is used as the loss function, as it has been shown to be ideal for classification tasks \cite{shore1981properties}.

As is standard in artificial neural networks, these algorithms undergo two stages, training and test. To train, validate and test all of our neural networks, we dedicated 70\% of the data-set to training, 15\% to validation, and 15\% to testing, whereas the number of epochs was always limited to 1000. In all cases, the networks were trained and tested with balanced data to avoid bias in identification, and the testing data was always excluded from the training stage. More specifically, we train our neural networks \cite{svozil1997introduction} from distorted modes collected by the CCD camera after the propagation through the fiber. The collection of RGB high-resolution images (1200$\times$1024 pixels) are converted into grayscale images by eliminating the hue and saturation information but retaining the luminance. To reduce the data dimension, a down-sampling process is performed on the resulting monochromatic images by averaging small clusters of 140$\times$140 pixels to form images of 9$\times$7 pixels. In this way, the feature vector is obtained by reorganizing the pixels of the resulting images as a column vector. At this point, it is important to stress that the proper choice of the feature vector can have a dramatic effect on the performance results. As shown in the main manuscript, our extreme reduction in the image resolution allows us to train neural networks in a short time with low computational resources while maintaining a high recognition rate. Once the NN has been trained, Bob can utilize its high efficiency to retrieve the message sent by Alice, even if the channel is under strain, with high confidence in both message security and integrity. In order to assess the performance of the neural networks, we compute the ratio of the sum of false negatives and false positives to the total number of input observations, the so-called accuracy. We have run all of our algorithms in a computer with an Intel Core i7–4710MQ CPU (@2.50GHz) and 32GB of RAM with MATLAB 2019a.

\bibliography{main}
\end{document}